\documentclass[referee]{raa}           

\usepackage{graphicx,times}
\usepackage{natbib}
\usepackage{amssymb,amsmath}
\usepackage{siunitx}
\bibpunct{(}{)}{;}{a}{}{,}

\newcommand{\sunmass}{{\rm M}_\odot}

\usepackage[pagebackref=true, colorlinks=true, linkcolor=blue, citecolor=blue, filecolor=blue, urlcolor=blue]{hyperref}

\begin{document}

   \title{The completeness of accreting neutron star binary candidates from Chinese Space Station Telescope}

 \volnopage{ {\bf 20XX} Vol.\ {\bf X} No. {\bf XX}, 000--000}
   \setcounter{page}{1}

   \author{Hao SHEN 
      \inst{1,2}
   \and Shun-Yi LAN
      \inst{1,2}
    \and Xiang-Cun MENG
      \inst{1,3}
   }
   \institute{Yunnan Observatories, Chinese Academy of Sciences, Kunming 650216, China;  \\{\it ~~~~shenhao@ynao.ac.cn, xiangcunmeng@ynao.ac.cn}\\
        \and
             University of Chinese Academy of Sciences, Beijing 100049, China\\
        \and
             International Centre of Supernovae, Yunnan Key Laboratory, Kunming 650216, China\\
\vs \no
   {\small Received 2024 June 17; accepted 2024 July 27}
}

\abstract{
Neutron star (NS) has many extreme physical conditions, and one may obtain some important informations about NS via accreting neutron star binary (ANSB) systems. The upcoming Chinese Space Station Telescope (CSST) provides an opportunity to search for a large sample of ANSB candidates. Our goal is to check the completeness of the potential ANSB samples from CSST data. In this paper, we generate some ANSBs and normal binaries under CSST photometric system by binary evolution and binary population synthesis method and use a machine learning method to train a classification model. Although the Precision ($94.56~ \%$) of our machine learning model is as high as before study, the Recall is only about $63.29~ \%$. The Precision/Recall is mainly determined by the mass transfer rate between the NSs and their companions. In addition, we also find that the completeness of ANSB samples from CSST photometric data by the machine learning method also depends on the companion mass and the age of the system. ANSB candidates with low initial mass companion star ($0.1~ \sunmass$ to $1~ \sunmass$) have a relatively high Precision ($94.94~ \%$) and high Recall ($86.32~ \%$), whereas ANSB candidates with higher initial mass companion star ($1.1~ \sunmass$ to $3~ \sunmass$) have similar Precision ($93.88~ \%$) and quite low Recall ($42.67~ \%$). Our results indicate that although the machine learning method may obtain a relative pure sample of ANSBs, a completeness correction is necessary for one to obtain a complete sample.
}

   \authorrunning{SHEN et al.}            
   \titlerunning{The completeness of ANSB candidates from CSST}  
   \maketitle

%
\section{Introduction}           
\label{sect:intro}
The concept `neutron star' (NS) was proposed by Lev Davidovich Landau in 1932, but until 1967 Antony Hewish and Jocelyn Bell Burnell detected the first known NS (also a pulsar, \citealt{1968Natur.217..709H}). Because NS possess many extreme physical parameters, it is a unique laboratory that allows us to study the properties of dense matter. Several decades later, many breakthroughs in physics and in astronomy are achieved about NSs, especially the first gravitational wave detection from the merging of two NSs (i.e., GW170817), which is `unprecedented joint gravitational and electromagnetic observation' and `marks the beginning of a new era of discovery' \citep{PhysRevLett.119.161101}. However, some crucial information of NS, such as structure, equation of states, birth environment and progenitor are still vague (\citealt{2012ARNPS..62..485L, 2013RPPh...76a6901O, 2015PASA...32...16S}).

Accreting neutron star binaries (ANSBs) are a sort of special binary systems: where NSs are accreting material from their normal companion stars through Roche lobe outflow (RLOF) or stellar wind accretion \citep{2002apa..book.....F}. The accreted materials may form a disk around the NSs and emit X-ray for releasing gravitational potential energy. Such binaries are called as `X-ray binaries'. Depending on the mass of companion stars, ANSBs are classified into three categories: low-mass X-ray binary (LMXB, companion star mass below $1~\sunmass$), intermediate-mass X-ray binary (IMXB, companion star mass between $1~\sunmass$ and $10~\sunmass$) and high-mass X-ray binary (HMXB, companion star mass higher than $10~\sunmass$) (\citealt{2018ASSL..457..185D}). Some ANSBs are persistent X-ray sources, while the others are transient X-ray sources (\citealt{2018SSRv..214...15D}).

As X-ray sources, ANSBs also emanate NUV/optical radiation. The NUV/optical radiation from ANSBs has several different physical origins: thermal radiation from accretion disk \citep{1973A&A....24..337S, 1995xrbi.nasa...58V, 2002apa..book.....F}, X-ray reprocess \citep{1976ApJ...208..534C, 1994A&A...290..133V, 2006MNRAS.371.1334R}, interaction between relativistic stellar wind of NS and inflow matter \citep{2000ApJ...541..849C}, synchrotron radiation in jet \citep{2002ApJ...573L..35C, 2005ApJ...624..295H, 2006MNRAS.371.1334R} and in hot accretion flow \citep{Veledina2012HotAF}, the companion star. In this parer, we mainly investigate the contribution from the accretion disk, the X-ray reprocess and the companion star. We will consider other mechanisms step by step to complete our model in the future.

ANSB plays an important role in the theory of binary evolution and the formation of millisecond pulsar (MSP). According to the recycling scenario, a slowly rotating neutron star in a binary system, which obtains angular momentum by accreting material from its companion star, can become a MSP (see \citealt{1991PhR...203....1B} for a review). During the accretion phase, the ANSB manifests itself as a X-ray source, while the binary hosts a radio MSP after mass transfer stops. The discovery of coherent pulsations in a transient LMXB SAX J1808.4-3658 \citep{1998Natur.394..344W} and transitional MSPs strongly support the recycling scenario \citep{2009Sci...324.1411A, 2013Natur.501..517P, 2014MNRAS.441.1825B}. So, ANSB may provide key information on the formation of MSP and binary evolutions (e.g. \citealt{1984ApJS...54..443P, 2019ApJ...887...48B}). However, the present small sample of ANSBs limit their roles to provide such enough information (e.g. \citealt{2003A&A...404..301R, 2007A&A...469..807L, 2023A&A...675A.199A}). A complete sample of ANSB is very important to constrain the formation of MSPs. The upcoming Chinese Space Station Telescope (CSST) could provide an opportunity to build such a sample. CSST is a ${\rm2\ m}$ space telescope and is planned to launch in a few years. The survey from CSST have seven photometric imaging bands covering {$255~\rm{nm}$ to $1000~\rm{nm}$}, with large field of view $\sim1\ \rm{deg}^2$, and a high-spatial resolution $\sim0.15''$ \citep{2018MNRAS.480.2178C, 2019ApJ...883..203G}. 

In a previous related work, \cite{2022RAA....22l5018L} produced some ANSBs and normal binary stars to investigate if machine learning can efficiently search for the ANSB candidates under the CSST photometric system. Their classification results indicate that machine learning can efficiently select out the ANSB candidates from the background normal stars. However, the ANSBs in their model were obtained by some binary evolutions, and then the parameter space of these ANSBs may not cover the whole range of the parameter space. They also did not check the completeness of ANSBs in CSST data. In this work, we will extend the parameter space of the generated ANSBs to check the completeness of the ANSB same obtained from the CSST photometric data, and to check which parameter mainly affects the completeness of ANSB sample.

The paper is organized as followings. In Section \ref{sect:method}, we describe our methods. We show the machine learning classification results in Section \ref{sect:results}. Discussions and conclusions are given in Section \ref{sect:discuss} and \ref{sect:conclusion}, respectively.

\section{Method}
\label{sect:method}

The main aim in the paper is to check Precision and Recall from the CSST photometric data by a machine learning method, and some basic methods are similar to these in \cite{2022RAA....22l5018L}. For example, following \cite{2022RAA....22l5018L}, we assume that the observed optical emission from an ANSB system are mainly contributed by its accretion disk and companion star. \cite{2022RAA....22l5018L} only consider main sequence stars as the companions, while the RGB (red giant branch) and AGB (asymptotic giant branch) stars are also included in this paper. The detailed methods in the paper are follows.

1. We used binary population synthesis to generate background stars ranging in age from $1~{\rm Myr}$ to $14~{\rm Gyr}$, with initial masses ranging from $0.5~ \sunmass$ to $10~ \sunmass$ and solar metallicity, in which all the systems are evolving in binaries, based on Hurley's rapid binary evolution code \citep{2000MNRAS.315..543H, 2002MNRAS.329..897H}. The basic assumptions for the binary population synthesis are similar to those in \cite{2009MNRAS.395.2103M} and \cite{2017MNRAS.469.4763M}.

2. We generate ANSB samples in our model by setting several independent variables (the mass of NS, the initial mass and the age of companion star, and the mass transfer rate). We assume that the NS in an ANSB system is a point mass, i.e., we neglect the radiation from surface of the NS and the spin of the NS. We generate the neutron stars with masses ranging from $1.4~ \sunmass$ to $2~ \sunmass$, by a Monte Carlo method. We focus on the cases that the companions fulfill their Roche lobes, i.e., we focus on the LMXBs and IMXBs, since HMXBs are mainly from wind accretion (\citealt{2002apa..book.....F}). Because there is a rough boundary of $3~ \sunmass$ for the properties of X-ray binary systems (\citealt{2023arXiv230612002Z}), based on models from \cite{2010ApJ...717..724G, 2015ApJ...812...40G}, the companions in the ANSBs are generated with discrete initial masses ranging from $0.1~ \sunmass$ to $3~ \sunmass$, and with an age evenly distributed during their whole lifes. The mass transfer rates are also randomly generated, from $\log{ \left(\frac{\mathop{M}\limits^{.}}{\sunmass/\rm{yr}} \right) }=-14$ to $\log{ \left(\frac{\mathop{M}\limits^{.}}{\sunmass/\rm{yr}} \right) }=-6$, where, the lower limit is set to be much lower than that from wind accretion rate, while the upper limit is set for typical thermal timescale mass transfer rate (\citealt{2010ApJ...717..724G}). Considering that the companions are fulling their Roche lobes, we set the binary separation $a$ following the equation in \cite{1983ApJ...268..368E}:
\begin{equation}\label{eq1}
a={\frac{{R_2}{({0.69}{q}^{2/3}+\ln{(1+{q}^{1/3})})}}{{0.49}{q}^{2/3}}},
\end{equation}
where $R_2$ is the radius of a companion star. The orbital inclination $\theta$ has a great influence on the observed optical flux from accretion disk, i.e., for given conditions, the flux is proportional to $\cos{\theta}$ (see \citealt{2002apa..book.....F}), we set $\theta$ as $0^{\circ}<\theta<90^{\circ}$, randomly.

3. The other methods to obtain the total magnitudes of ANSBs, including the systematic error of CSST photometric system are the same as those in \cite{2022RAA....22l5018L}. In paticular, the radiation from the disk includes those from both multi-color disk and irradiate accretion disk, as in \cite{2022RAA....22l5018L}. We also use the same machine learning process as in \cite{2022RAA....22l5018L} to compare our results with these in \cite{2022RAA....22l5018L}.

\begin{figure}[t]
    \centering
    \includegraphics[width=1\linewidth]{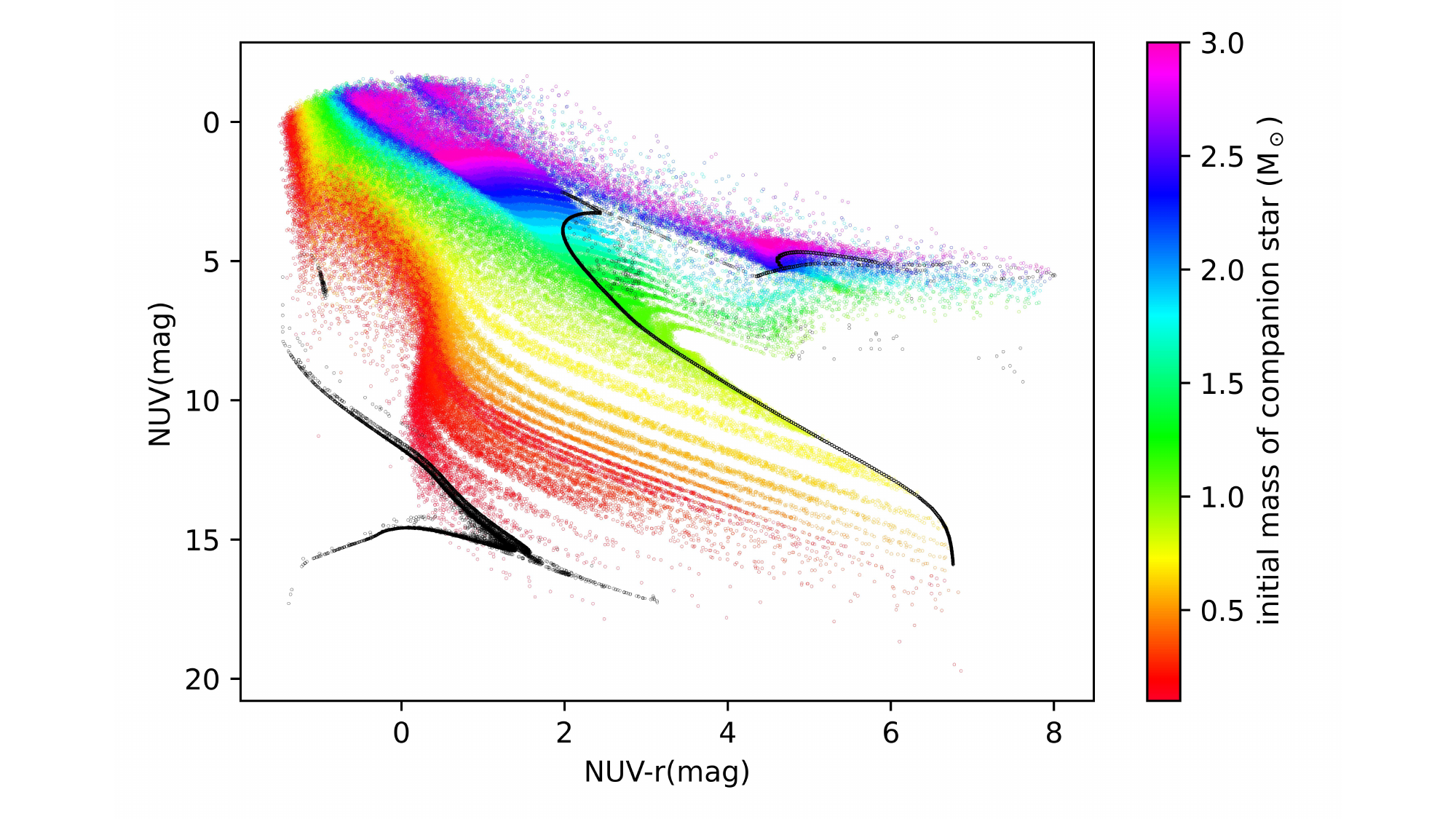}
    \caption{The CMD for ANSBs (colorful dots), where different colors represent different initial masses of companion stars. For comparison, single stars with different masses and an age of $1~{\rm Gyr}$ are also shown.}
    \label{fig1}
\end{figure}

In Fig~\ref{fig1}, we show all the ANSBs we generated (colorful dots) in the color-magnitude diagram (CMD), where different colors represent different initial masses of companion stars. For comparison, some background single stars with $1~{\rm Gyr}$ age (black dots) are also shown. The ANSBs cover a large range in the CMD because the systems have different NS masses, different companion masses, different ages and different mass transfer rates. The positions of ANSBs in CMD are strongly dependent on mass transfer rates. For a given companion, ANSBs with higher mass transfer rates will appear brighter and bluer, while ANSBs with lower mass transfer rates are located close to normal single stars in the CMD, as shown by the discrete tapes for low mass stars. In Fig~\ref{fig1}, there is a clear upper boundary for the brightness of ANSBs, which is from our treatment that there is an upper limit of mass transfer rate of $10^{-6}~\sunmass/\rm{yr}$.

\section{Results}
\label{sect:results} 
There are several metrics about classification results in machine learning, among them, Precision = $\frac{\rm{TP}}{\rm{TP+FP}}$ represents the proportion of examples that are divided into positive cases that are actually positive cases, Recall = $\frac{\rm{TP}}{\rm{TP+FN}}$ represents the proportion of all positive cases that are correctly classified, and measures the model's ability to recognize positive cases. Recall may represent the completeness of recognized ANSBs, from observational data. Here, TP (True Positive) means a positive example that is correctly predicted, i.e., the true value of the data is positive and the predicted value is also positive. TN (True Negative) means a negative example that is correctly predicted. FP (False Positive) means a negative example that was incorrectly predicted. FN (False Negative) means a positive example that was incorrectly predicted.

For the whole machine learning sample, the Precision is $94.56~\%$, similar to that in \cite{2022RAA....22l5018L}. In other words, almost all of the ANSB candidates identified from the CSST photometric data may be real ANSBs. However, the Recall is only $63.29~\%$, which means that many ANSBs may be missed for a machine learning method. Actually, as well konwn, there is a competition between Precision and Recall for a machine learning model, in which a key threshold between 0 and 1 is designed to balance the competition. In our model, the threshold is set to be 0.5. We did a test for the threshold of 0.3 and found that Precision is decreased to be $89.42~ \%$ and Recall is increased to be $66.67~ \%$, as expected. Table~\ref{tab:1} shows the classification results of our model.

\begin{table*}[t]
    \centering
    \caption{Total classification results}
    \label{tab:1}
    \begin{tabular}{cccc}
    \hline
        \textbf{} & \textbf{Correctly classified} & \textbf{Falsely classified} & \textbf{All} \\ 
        \hline
        ANSBs & 49822 (TP) & 28888 (FN) & 78710\\
        Background stars & 31539025 (TN) & 2863 (FP) & 31541888\\
        All & 31588847 & 31751 & 31620598\\
        \hline
    \end{tabular}
\end{table*}

\begin{figure}[]
    \centering
    \includegraphics[width=1\linewidth]{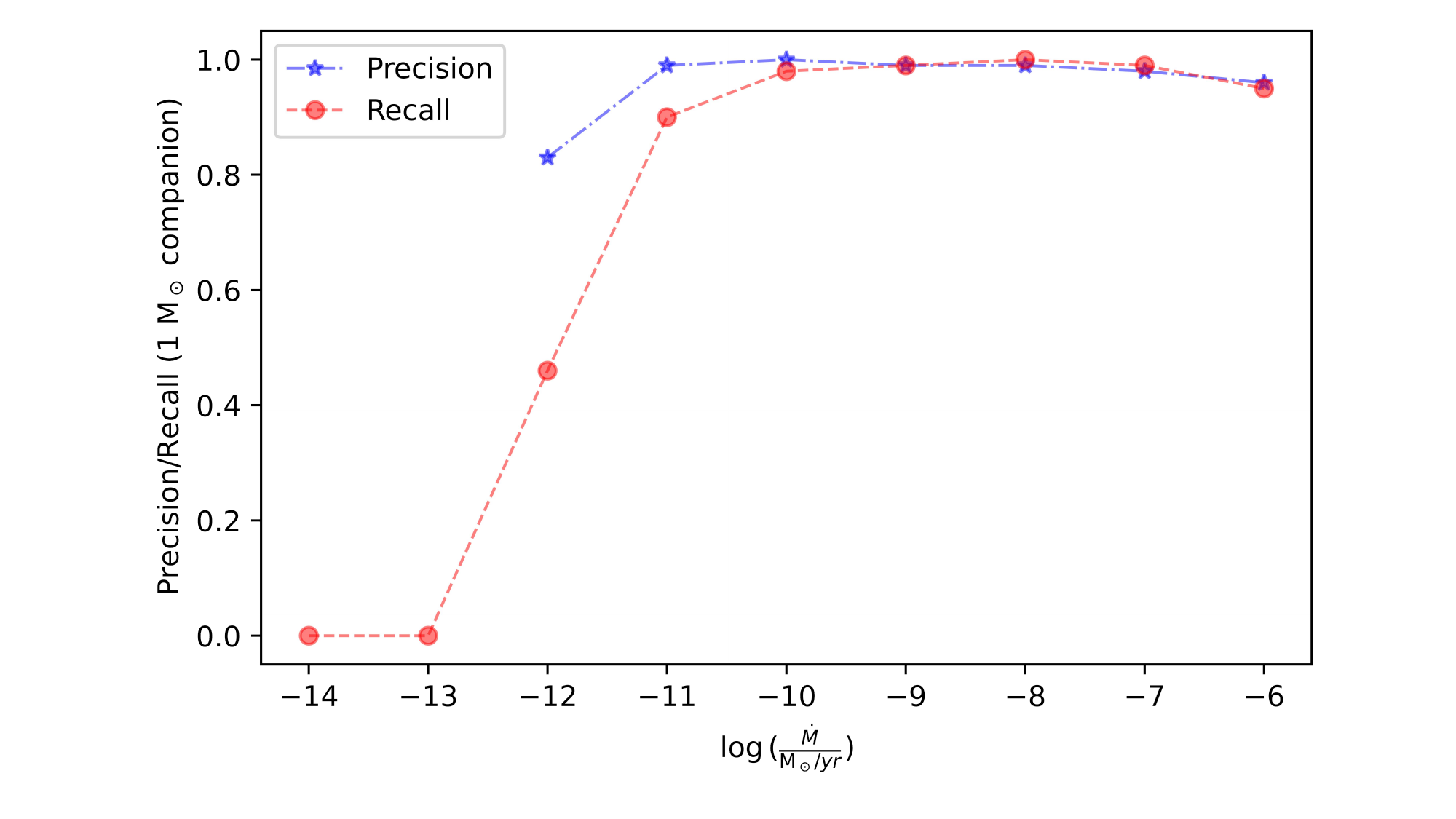}
    \includegraphics[width=1\linewidth]{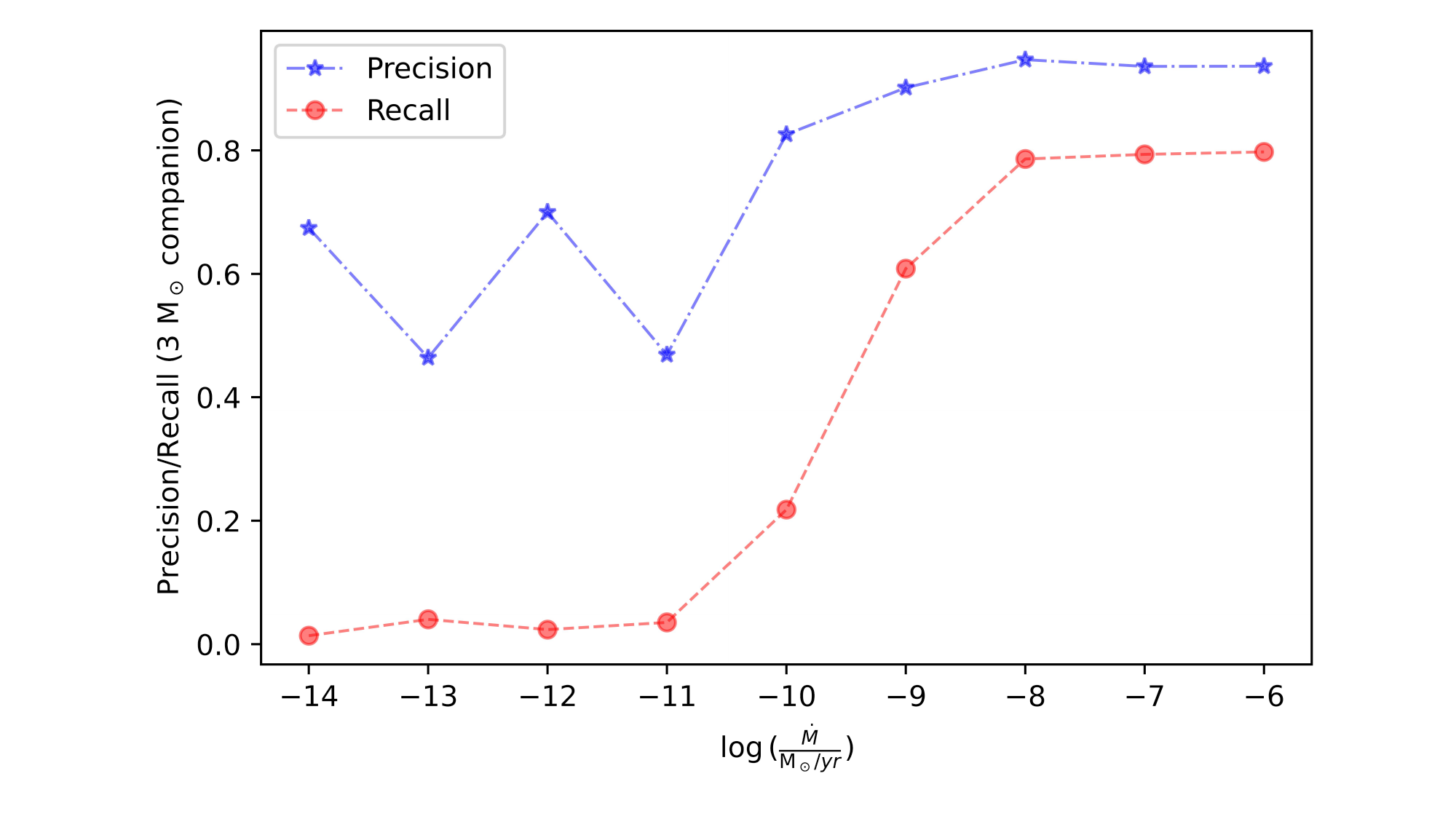}
    \caption{Precision and Recall of ANSB systems as a function of mass transfer rates. The pentagrams represent Precision, and the circles represent Recall. Upper panel is the case for the companion of $1~ \sunmass$, while the lower panel is for the case of the compaions of $3~ \sunmass$.}
    \label{fig2}
\end{figure}

As we shown in Fig~\ref{fig1}, where an ANSB is recognized by the machine learning method could be mainly determined by the mass transfer rate between the NSs and their companions. In Fig~\ref{fig2} we show the Precision and Recall of ANSBs as a function of the mass transfer rate, where other parameters are randomly distributed within the range of values we set. For ANSBs with a given companion star, the higher the mass transfer rate, the higher the Precision and the Recall. In the upper panel of Fig~\ref{fig2}, we show the Precision and Recall of ANSB systems with companion stars of $1~\sunmass$ as a function of the mass transfer rate. For ANSB systems with the companion stars of $1~\sunmass$, our model cannot identify samples with mass transfer rates of less than $\log{ \left(\frac{\mathop{M}\limits^{.}}{\sunmass/\rm{yr}} \right) }=-12$. While our model is very good at identifying samples with mass transfer rates higher than $\log{ \left(\frac{\mathop{M}\limits^{.}}{\sunmass/\rm{yr}} \right) }=-11$, these samples have Precision and Recall close to 1. For comparison, we show in lower panel of Fig~\ref{fig2} the Precision and Recall of ANSB systems with the companion stars of $3~\sunmass$ as a function of mass transfer rate. Our model cannot efficiently identify samples with a mass transfer rate of less than $\log{ \left(\frac{\mathop{M}\limits^{.}}{\sunmass/\rm{yr}} \right) }=-9$. Samples with mass transfer rates higher than $\log{ \left(\frac{\mathop{M}\limits^{.}}{\sunmass/\rm{yr}} \right) }=-9$ can well be identified, but about one fifth of their completeness is missing. In other words, for a given companion, Precision/Recall reaches almost a constant value. This is from the fact that when mass transfer rate is high enough, the radiation from the accretion disk of an ANSB dominate the observable optical flux, i.e., the radiation from companion star is negligible, compared with that from the accretion disk. The results here indicate that the CSST photometric system could be sensitive to those ANSBs with thermal-timescale mass transfer.

\begin{figure}[]
    \centering
    \includegraphics[width=1\linewidth]{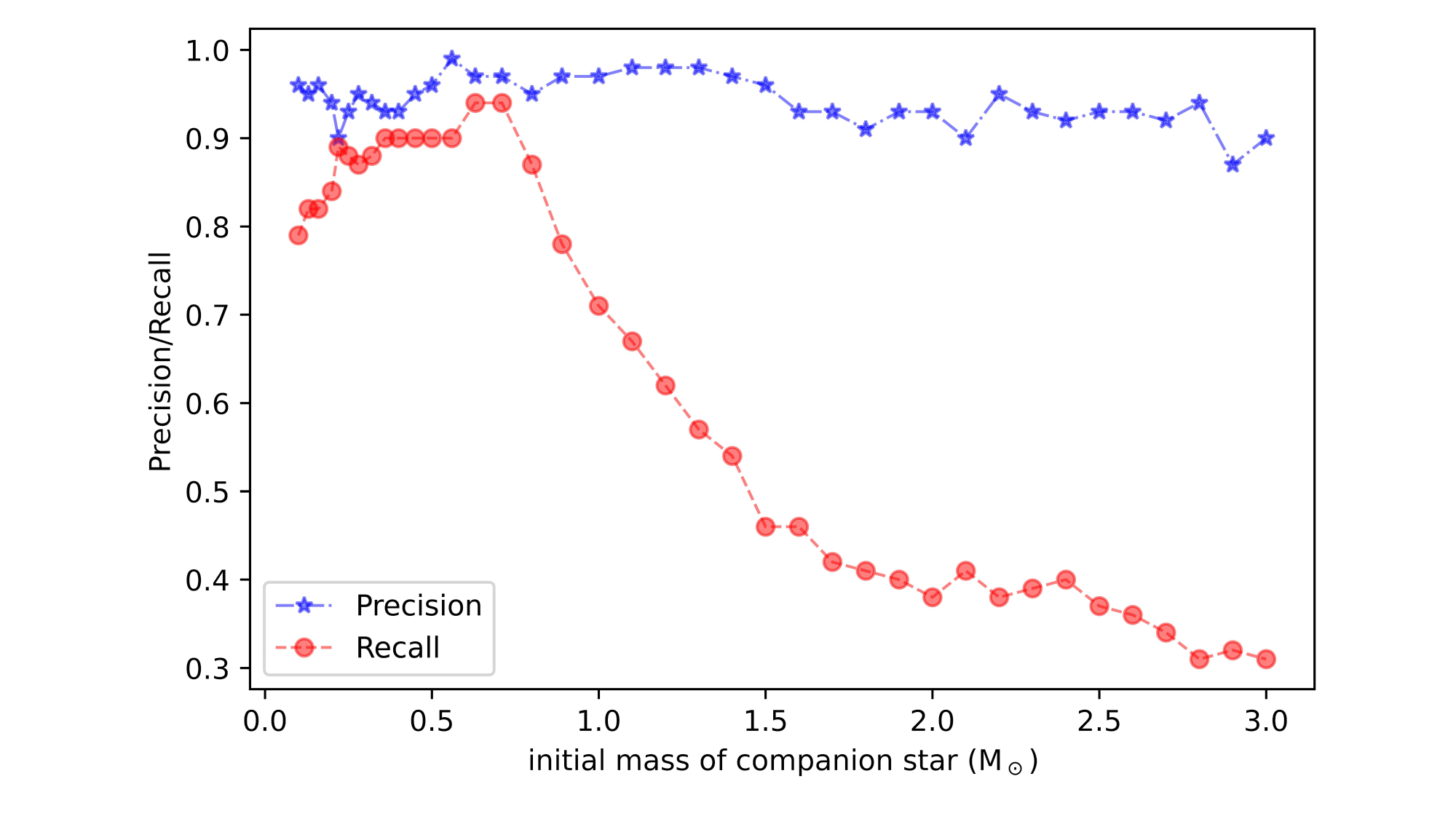}
    \caption{Precision and Recall of ANSB systems as a function of the initial masses of companion stars. The pentagrams represent Precision, and the circles represent Recall.}
    \label{fig3}
\end{figure}

As shown in Fig~\ref{fig2}, our model can easily identify ANSB systems with high mass transfer rates, but the completeness is affected by the initial masses of companion stars. In Fig~\ref{fig3}, we show the Precision and Recall of ANSBs as a function of the initial mass of companion star, where other parameters are randomly distributed within the range of values we set. During the whole interval of the companion mass, the Precision is almost a constant of about $94~\%$, with a slightly higher value for the stars less than $1~\sunmass$ than more massive stars. As discussed by \cite{2022RAA....22l5018L}, the high Prcision is derived from the fact that the flux of a normal binary is roughly to be the superposition of two blackbody spectra, but the flux from our ANSB model is roughly the superposition of an accretion disk and a blackbody spectrum. Such a huge flux difference decides the high Precision. Moreover, a more massive companion star makes its spectrum to move to a state more similar to an accretion disk as shown in \cite{2022RAA....22l5018L}, which results in a slightly decrease of the Precision with the companion mass. However, the Recall is highly dependent on the companion mass. For the cases with the companion of $\textless ~1~\sunmass$, the Recall is significantly higher than that from the cases with the companion of $\textgreater ~1~\sunmass$ ($86.32~\%$ vs $42.67~\%$). In paticular, the Recall decreases with the increase of companion mass for the cases with companions of $\textgreater ~0.71~\sunmass$. This is due to the fact that the greater the initial mass of companion star, the brighter the companion star becomes, and the flux from the companion star may cover the flux from the accretion disk.

\begin{figure}[]
    \centering
    \includegraphics[width=1\linewidth]{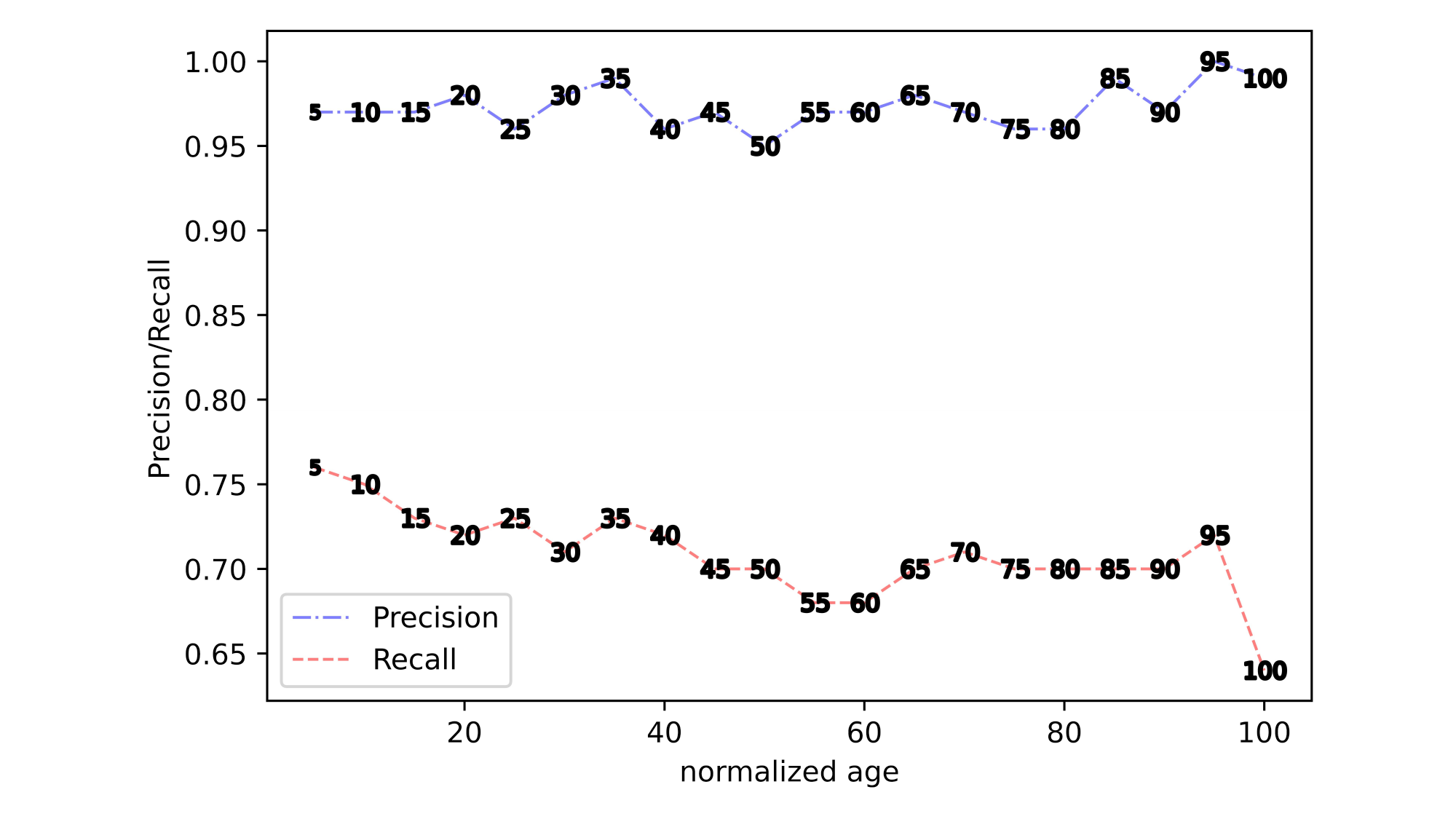}
    \includegraphics[width=1\linewidth]{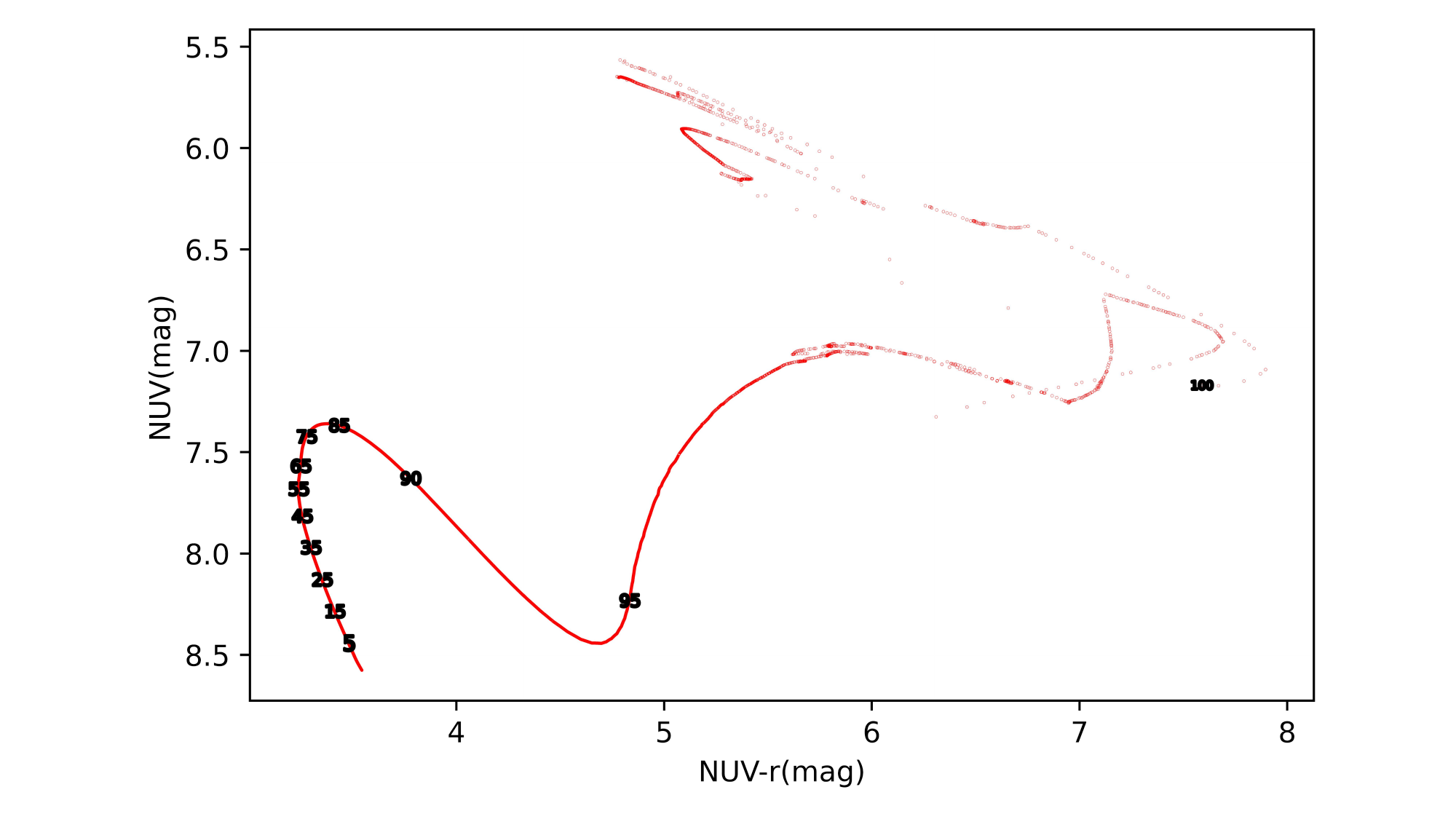}
    \caption{Upper panel: Precision and Recall of ANSB systems (with $1~\sunmass$ initial mass companion stars) as a function of age. The dot dashed line represent Precision, and the dashed line represent Recall. We set the lifetime of companion stars from age 0 to age 100 (normalized age) and label Precision/Recall with the corresponding ages; Lower panel: The evolutionary track of stars with initial mass of $1~ \sunmass$ (ZAMS to terminal AGB) in a CMD. To facilitate comparison with the upper panel, we have marked the ages of some dots.}
    \label{fig4}
\end{figure}

The age of ANSB, which is determined by the companion age, could also affect the Precision and/or Recall. The upper panel of Fig~\ref{fig4} presents Precision and Recall of ANSBs as a function of the age of ANSB system, where their companion stars have the same initial mass ($1~ \sunmass$), other parameters are randomly distributed within the range of values we set. For the whole age stage, we have an overall result, Precision $= 97.32~ \%$ and Recall $= 70.95~ \%$. For different ages of ANSB systems, Precision is kept at high value ($\textgreater~95~\%$) with small fluctuation. This is because the radiation of a normal binary system comes from two blackbodies, while the radiation of an ANSB system comes from a blackbody and an accretion disk. The recall remains in the range of about 0.7 to 0.75 for $95~\%$ of the time and drops below 0.7 in the last $5~\%$ of the time due to the companion star being in main sequence stage and Hertzsprung gap for $95~\%$ of the time, resulting in minimal changes in brightness and color (see lower panel).

\begin{figure}[]
    \centering
    \includegraphics[width=1\linewidth]{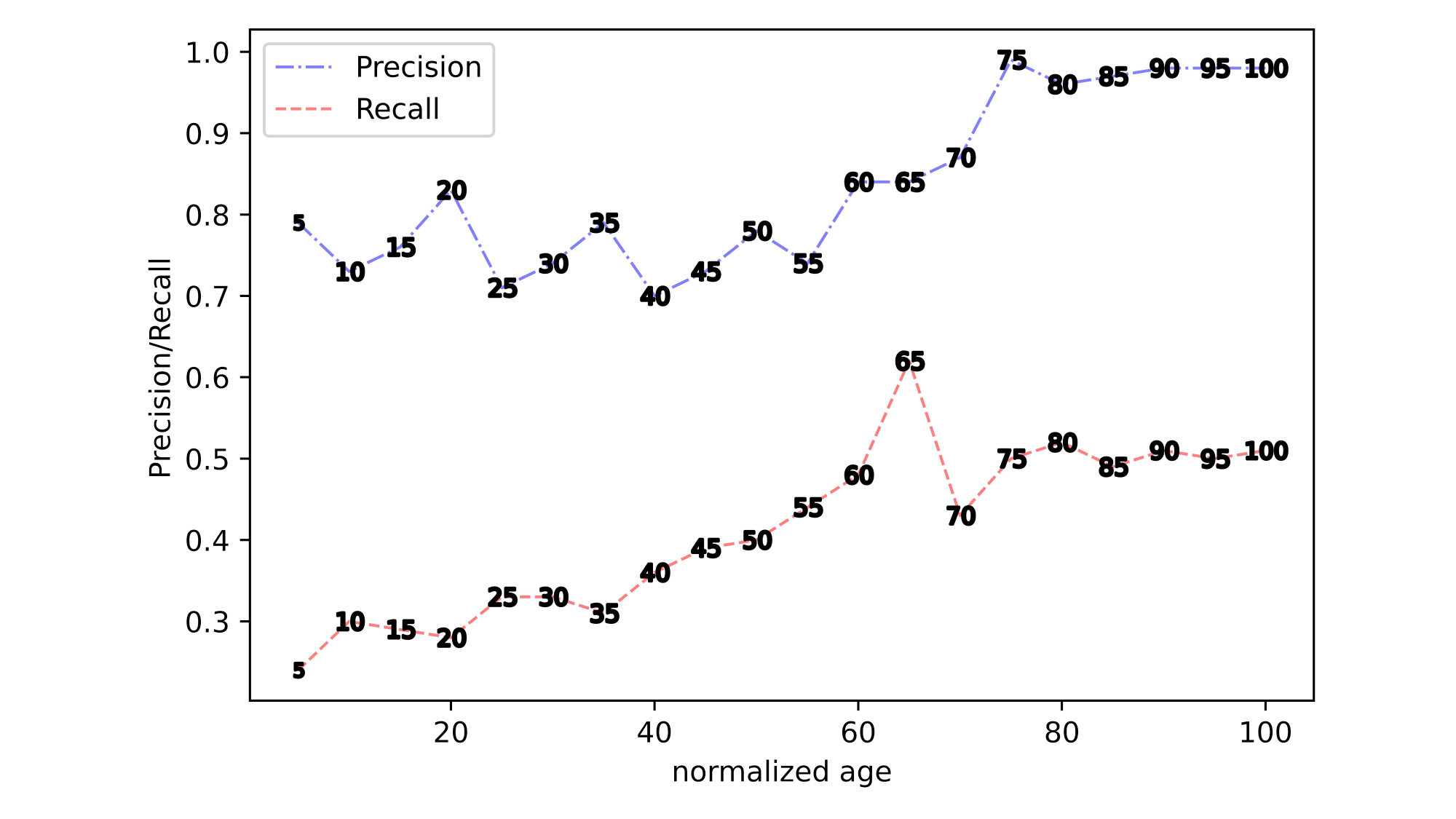}
    \includegraphics[width=1\linewidth]{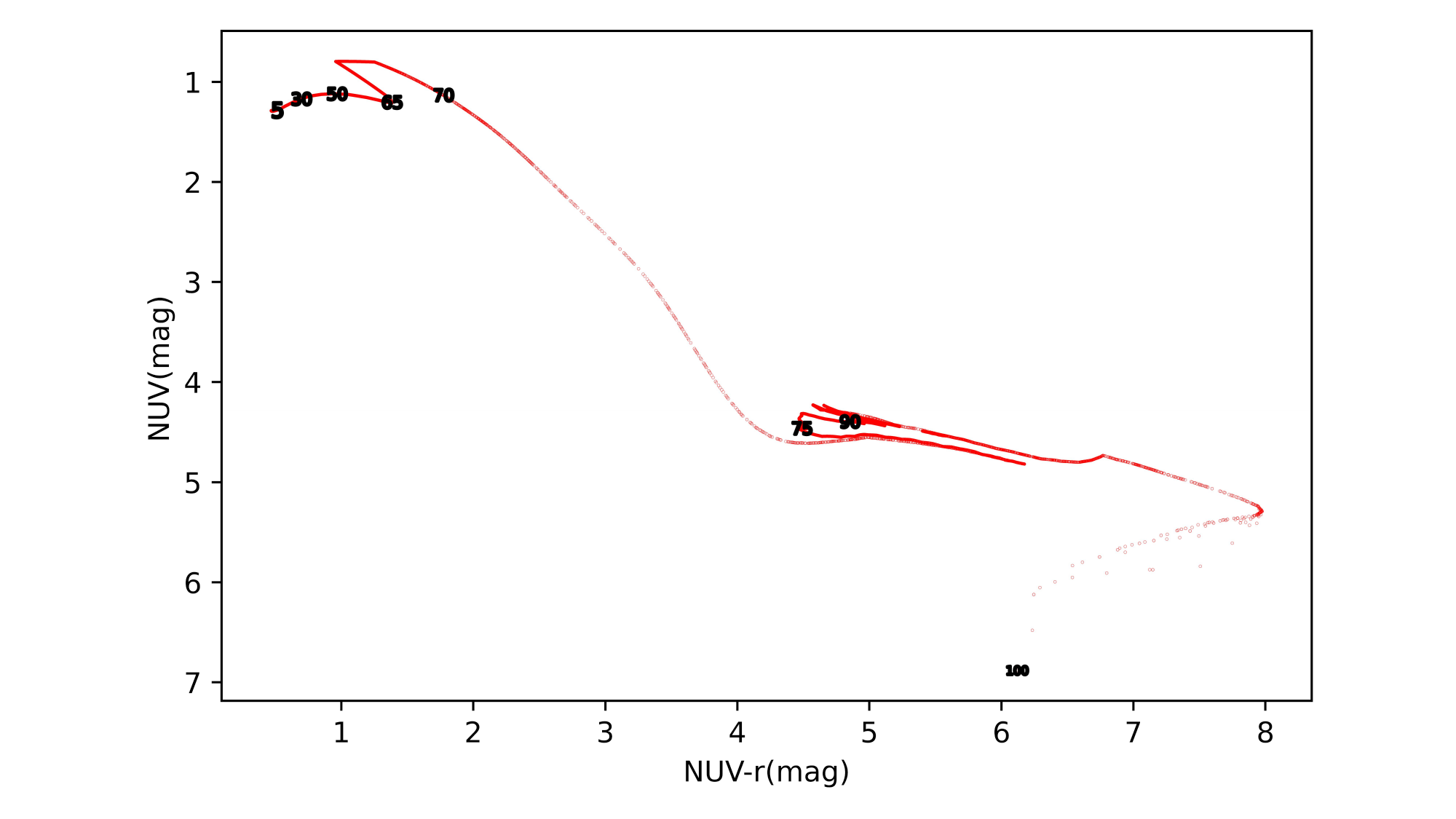}
    \caption{Upper panel: Precision and Recall of ANSB systems (with $3~\sunmass$ initial mass companion stars) as a function of age. The dot dashed line represent Precision, and the dashed line represent Recall, We set the lifetime of companion stars from age 0 to age 100 (normalized age) and label Precision/Recall with the corresponding ages; Lower panel: The evolutionary track of stars with initial mass of $3~ \sunmass$ (ZAMS to terminal AGB) in a CMD. To facilitate comparison with the upper panel, we have marked the ages of some dots.}
    \label{fig5}
\end{figure}

The age of a star heavily depends on its initial mass, so we must consider the influence of initial mass. Similar to Fig~\ref{fig4}, the upper panel of Fig~\ref{fig5} illustrates the Precision and Recall of ANSBs with ages, while their companion stars have an initial mass of $3~ \sunmass$. For the whole age stage, we have an overall result of Precision $= 83.58~ \%$ and Recall $= 41.47~ \%$. The Precision initially fluctuates in the range of 0.7 to 0.8 (age 0 to age 70), and then stabilizes close to a value of 1 (age 70 to age 100). This behavior is attributed to the high surface temperature of a star with an initial mass of $3~ \sunmass$ during its main sequence stage (age 0 to age 70, see lower panel), because ANSB systems with high temperature companion stars are relatively easily identified to be hot background stars. While Recall rises first and then stays around 0.5. This is because from age 0 to age 65, as the surface temperature of the companion star decreases, the radiation from ANSB system can exhibit both components from the accretion disk and the companion star, rather than just a single hot component. From age 65 to age 70, the brightness of the companion star increases, and the radiation from the companion star may partially cover the radiation from the accretion disk, leading to the decrease in Recall, i.e., the systems may be easily indentified as a single star by our machine learning model. Eventually, from age 70 to age 100, the companion star became a giant, and Recall remains stable.

\section{Discussion}
\label{sect:discuss}

In this paper, we check how the completeness of potential ANSB candidates from future CSST photometric data depends on the properties of ANSB, where three parameters are investigated, i.e., mass transfer rates, initial masses and the age of companions.

First, we found that the mass transfer rate has a decisive effect on the recognition results of our model. For ANSB systems with low mass transfer rates, our model is almost unable to identify them. For ANSB systems with sufficiently high mass transfer rates, our model can identify them well, and the recognition results have both high Precision and high Recall. This means that there is a high probability that X-ray emission can be observed in the ANSB systems identified by our model. Mass transfer rate is the dominant parameter to affect the Recall from a machine learning method, because the value directly determines the flux from the accretion disk in an ANSB system \citep{2022RAA....22l5018L}. However, the initial mass of companion star also has an impact on the identification results. A higher mass transfer rate `threshold' is required to produce good identification results when the initial mass of companion star is larger. Conversely, a lower upper limit of Recall is associated with a greater initial mass of companion star. The results here indicate that the CSST photometric system could be quite sensitive to the ANSB systems that a mass transfer occurs within thermal timescale.

In addition, we find that the initial mass of companion star has little effect on Precision and has a significant effect on Recall. For different initial masses of companion star, Precision is almost a constant (94~\%). As discussed by \cite{2022RAA....22l5018L}, the high Precision is due to the fact that the flux from normal binary star is superposition of two blackbody spectra, while the flux from ANSB is superposition of an accretion disk spectrum and a blackbody spectrum. However, the initial mass of the companion star has a significant effect on Recall. This is because the initial mass determines the brightness of companion star, and the brightness of companion star affects the flux contribution from companion star in ANSB system. Considering the effect of the mass transfer rate, the systems identified from the CSST photometric system are more likely to be LMXBs, which could be observed by X-ray observations.

We also explored the effect of the age of companion star on Precision and Recall, i.e., the more massive the companion, the more significant of the age effect on the Precision and/or Recall. This is due to the higher the initial mass of companion star, the greater the luminosity.

Finally, we should point out that the ANSBs in our model are produced based on some simple assumptions. For example, we only consider the radiation of ANSB from the accretion disk and the companion star; while the corona of the NS and the heating of the companion star are ignored (e.g. \citealt{1997MNRAS.285..673H, 2005ApJ...635.1203M, 2016ApJ...828....7R}). Some physical processes that we neglected could become important in some special ANSBs, such as, jet \citep{2006MNRAS.371.1334R, 2007MNRAS.379.1108R}. We also do not consider the effect of emission lines from the disk, which could affect the accuracy of the classification model (a double peak H$_\alpha$ emission line maybe support the existence of accretion disk, \citealt{2009ApJ...703.2017W}). In the future, we will add these effects step by step. Maybe the slitless spectroscopy module of CSST might be helpful to improve our model.

\section{Conclusions}
\label{sect:conclusion}

In this paper, we produced some ANSBs and background stars to investigate whether or not the completeness of ANSB candidates in CSST data depends on the properties of the potential ANSB systems. We obtain a total result of Precision $=94.56~ \%$ and Recall $=63.29~ \%$. We found that the mass transfer rate had a decisive influence on the identification results of our model. For ANSB systems with high mass transfer rates, our model is able to obtain good identification results (high Precision and high Recall). The initial mass and age of the companion star also have an impact on the identification results. Our results indicate that the ANSB systems with less massive companions and thermal timescale mass transfer rates are more likely to be identified by the CSST photometric system and would have a better completeness.

\normalem
\begin{acknowledgements}
We thank Zheng-Wei Liu, Hong-Wei Ge, Jun-Feng Cui, Jing-Xiao Luo, Li-Fu Zhang and Li-Jun Zhang for their help. This work is supported by the National Natural Science Foundation of China (Nos. 12288102 and 12333008), and the National Key R\&D Program of China (No. 2021YFA1600403). X.M. acknowledges support from the International Centre of Supernovae, Yunnan Key Laboratory (No. 202302AN360001), the Yunnan Revitalization Talent Support Program-Science \& Technology Champion Project (No. 202305AB350003), the Yunnan Fundamental Research Projects (No. 202401BC070007 and 202201BC070003), and the science research grants from the China Manned Space Project.
\end{acknowledgements}
  
\bibliographystyle{raa}
\bibliography{bibtex}

\end{document}